\documentclass[lettersize,journal]{IEEEtran}
\ifCLASSINFOpdf
\else
\fi
\usepackage{float}
\usepackage{setspace}
\usepackage{graphicx}
\usepackage{subfigure}
\usepackage{multicol}
\usepackage{algorithm,algorithmic}
\usepackage{amsmath}
\usepackage{mathrsfs}
\usepackage{amssymb, amsthm}
\usepackage{bm}
\usepackage{cite}
\usepackage{enumerate}
\usepackage{diagbox}
\usepackage{color}
\usepackage{setspace}
\usepackage{multirow}
\usepackage{enumitem}
\usepackage{caption}

\newcommand{\cK}{\mathcal{K}}

\newcommand{\cI}{\mathcal{I}}

\newcommand{\bbE}{\mathbb{E}}

\newcommand{\br}{\boldsymbol r}

\newcommand{\bP}{\boldsymbol P}

\allowdisplaybreaks[4]

\usepackage{amsmath}

\usepackage{multirow}

\begin{document}
%
\title{ Content-Aware RSMA-Enabled Pinching-Antenna Systems for Latency Optimization in 6G Networks
}
%
%
%

\author{Yu Hua,~Yaru Fu,~Yalin Liu,~Zheng Shi,~\IEEEmembership{Member,~IEEE}, Kevin Hung,~\IEEEmembership{Senior Member,~IEEE}

\thanks{  This work was supported in part by the Team-based Research Fund under Reference No. TBRF/2024/1.10, in part by the Research Matching Grant under Reference No. CP/2025/1.1, in part by the Hong Kong Metropolitan University Research \& Development Project under Reference No. RD/2023/2.22, and in part by the grant from the Research Grants Council of the Hong Kong Special Administrative Region, China, under project No. UGC/FDS16/E15/24, and in part by Wuxi University Research Starat-up Fund For High-level Talents 2025r030. \textit{(Corresponding author: Yaru Fu)}

Y. Hua is with the School of Electronic Information and Engineering, Wuxi University, Wuxi, Jiangsu 214105, China and Jiangsu Province Engineering Research Center of Integrated Circuit Reliability Technology and Testing System, Wuxi, Jiangsu 214105, China(Email: 860474@cwxu.edu.cn).

Y. Fu, Y. Liu, and K. Hung are with the School of Science and Technology,  Hong Kong Metropolitan University, Hong Kong, 999077, China (E-mail: yfu@hkmu.edu.hk; ylliu@hkmu.edu.hk, khung@hkmu.edu.hk).

Z. Shi is with the School of Intelligent Systems Science and Engineering, Jinan University, Zhuhai 519070, China (e-mail:zhengshi@jnu.edu.cn).}

}

\maketitle

\begin{abstract}
The Pinching Antenna System (PAS) has emerged as a promising technology to dynamically reconfigure wireless propagation environments in 6G networks.
By activating radiating elements at arbitrary positions along a dielectric waveguide, PAS can establish strong line-of-sight (LoS) links with users, significantly enhancing channel gain and deployment flexibility, particularly in high-frequency bands susceptible to severe path loss.
To further improve multi-user performance, this paper introduces a novel content-aware transmission framework that integrates PAS with rate-splitting multiple access (RSMA).
Unlike conventional RSMA, the proposed  RSMA scheme enables users requesting the same content to share a unified private stream, thereby mitigating inter-user interference and reducing power fragmentation.
We formulate a joint optimization problem aimed at minimizing the average system latency by dynamically adapting both antenna positioning and RSMA parameters according to channel conditions and user requests.
A Content-Aware RSMA and Pinching-antenna Joint Optimization (CARP-JO) algorithm is developed, which decomposes the non-convex problem into tractable subproblems solved via bisection search, convex programming, and golden-section search.
Simulation results demonstrate that the proposed CARP-JO  scheme consistently outperforms Traditional RSMA, NOMA, and Fixed-antenna systems across diverse network scenarios in terms of latency, underscoring the effectiveness of co-designing physical-layer reconfigurability with intelligent communication strategies.
\end{abstract}


%
\IEEEpeerreviewmaketitle


\section{Introduction} \label{Introduction}
The evolution towards sixth-generation (6G) wireless networks introduces stringent demands for massive connectivity, high capacity, and ultra-low latency, posing significant challenges to system spectral efficiency and connection reliability. In high-frequency bands, such as millimeter-wave and terahertz, communication links are highly dependent on line-of-sight (LoS) conditions. However, obstacles in practical environments often lead to non-LoS scenarios and severe path loss. To address this, Pinching Antenna System (PAS) has emerged as a promising solution by dynamically reconfiguring wireless propagation environments. Specifically, PAS utilizes dielectric waveguides to guide electromagnetic waves, and pinching antenna (PA) can be activated at arbitrary points along the waveguide to establish strong LoS links with users, thereby significantly enhancing channel strength and flexibility of deployment \cite{PA1,PA2,PA14}.
Concurrently, at the multiple access layer, Rate-Splitting Multiple Access (RSMA) has emerged as a powerful and robust framework that generalizes and outperforms conventional schemes, such as Orthogonal Multiple Access (OMA) and Non-Orthogonal Multiple Access (NOMA) \cite{RSMApro1,RSMApro2}. By splitting user messages into common and private parts, RSMA efficiently manages multi-user interference, offering a superior trade-off between performance and receiver complexity.

Although the individual merits of PAS and RSMA are evident, existing research on their integration remains content-agnostic, which is a critical oversight. In modern networks, especially in video streaming and content distribution scenarios, the transmitted data is not merely anonymous bits but structured ``content", distinct files such as videos, songs, or software updates from a finite library. A key characteristic of content delivery is multicast; multiple users often request the same popular file simultaneously. However, conventional RSMA, even in such scenarios, generates a distinct private stream for each user, irrespective of whether they request the same content. This leads to the inefficient transmission of duplicate streams, creating unnecessary inter-user interference and fragmenting the limited transmit power across multiple streams for the same data. This power fragmentation and increased interference directly result in suboptimal data rates and, consequently, higher latency, which is the very metric 6G aims to minimize.

Therefore, the core motivation of this work is to bridge this gap by introducing content-awareness into the PAS-RSMA co-design. We propose that the system recognizes when users request the same content and adapts its transmission strategy accordingly. This paradigm shift, from user-aware to content-aware transmission, unlocks significant potential for latency reduction. Furthermore, the physical reconfigurability of PAS provides a unique spatial degree of freedom that can be jointly optimized with this content-aware RSMA strategy to further enhance the channel conditions for both common and private streams. The joint optimization of these cross-layer parameters, spanning the physical layer (antenna position) and the link layer (content-aware RSMA resource allocation), specifically for minimizing latency in a content-aware network, constitutes the central problem addressed in this paper.
\subsection{Literature Review} 

Existing research on PAS has primarily focused on their fundamental capabilities and performance in single-user or basic multi-user scenarios. The seminal work in \cite{PA1} introduced the core PAS concept, demonstrating that applying a plastic pinch to a dielectric waveguide enables efficient radio wave radiation in indoor environments. This was further expanded in \cite{PA2}, which provided a comprehensive overview of PAS principles, applications, and challenges, with particular emphasis on its potential to improve LoS communication and coverage extension. Authors in \cite{PA14} developed a comprehensive 3D modeling and theoretical framework for PAS in indoor immersive communications, providing both performance analysis and practical deployment guidelines for 6G applications.
Performance evaluations in \cite{PA3} established PAS's superiority over conventional fixed-antenna systems in mitigating path loss, supported by detailed analyses of outage probability and average data rate. Building on this, authors in \cite{PA4} investigated rate maximization in a downlink PAS through optimized antenna positioning to minimize path loss while ensuring constructive signal interference.

Recent studies have explored more advanced multi-user configurations and application scenarios. In \cite{PA7}, a matching algorithm was developed for joint waveguide assignment and antenna activation, strategically establishing LoS/NLoS links to enhance signals and suppress interference, thus significantly improving the sum rate. The authors in \cite{PA8} revealed that LoS blockage can actually enhance PAS performance in multi-user scenarios by suppressing co-channel interference. For multi-waveguide configurations, the authors in  \cite{PA9} proposed a fractional programming approach to optimize antenna locations, achieving significant gains in uplink sum of rates while maintaining robustness to waveguide losses.

The application scope of PAS has expanded to encompass a diverse range of domains. In \cite{PA10}, a novel wireless sensing architecture integrating PAS with leaky coaxial cables was proposed, where a particle swarm optimization (PSO)-based algorithm jointly optimizes the transmit waveform and antenna positions to minimize the Cramér-Rao bound (CRB), demonstrating substantial improvements in sensing accuracy.  Meanwhile, \cite{PA12} applied maximum entropy reinforcement learning to joint antenna positioning and power optimization in PAS-enabled integrated sensing and communication (ISAC) systems, demonstrating superior performance in balancing communication rates and sensing requirements.

Collectively, these studies underscore the capability of PAS to enhance channel conditions through dynamic antenna placement. However, most existing works focus on single-user scenarios or simple orthogonal multiple access (OMA) schemes, leaving key challenges such as multi-user interference and resource allocation largely unaddressed.
To address these challenges, advanced multiple access techniques, including non-orthogonal multiple access (NOMA) and rate-splitting multiple access (RSMA), have been integrated into PAS designs. NOMA, in particular, leverages power-domain multiplexing to serve multiple users within the same resource block, naturally aligning with the signal superposition characteristics inherent in waveguide-based PAS architectures.
For instance, the work in \cite{PA5} investigated sum-rate maximization in an uplink NOMA-assisted PAS, establishing its superior spectral efficiency compared to OMA schemes. Extending this line of research, \cite{PA6} explored a NOMA-based PAS with discrete antenna activation, developing low-complexity algorithmic solutions for sum-rate optimization. Further advancing system efficiency, \cite{PA13} introduced a convergent iterative algorithm for multi-waveguide NOMA-assisted PAS that minimizes total transmit power while revealing an oscillatory decay relationship between the optimal power and waveguide spacing. For wireless powered communications, the authors in \cite{PA11} investigated PAS-assisted networks with both TDMA and NOMA protocols, developing high-performance element-wise (EW) and low-complexity stochastic parameter differential evolution (SPDE) algorithms that achieve optimal performance with a PA power distribution ratio of 0.55-0.6.

However, the practical implementation of NOMA imposes a significant computational burden, particularly on users with strong channel conditions. These users are required to sequentially decode and cancel the signals intended for all weaker users via successive interference cancellation (SIC) before accessing their own information. This process introduces considerable processing complexity and latency, which becomes prohibitive as the number of users increases \cite{NOMAfuza}. To mitigate these inherent limitations of NOMA, RSMA has been advanced as a more flexible and robust alternative \cite{RS6,RS7,RS8}. In a downlink RSMA system, each user's message is split into a common part and a private part. The base station consolidates all common parts into a single common stream, while encoding each private part into an individual stream. All streams are then superimposed for transmission. At the receiver, each user first decodes the common stream, subtracts it via SIC, and then proceeds to decode its designated private stream while treating other private streams as noise. A key advantage of RSMA lies in its ability to manage complexity by adjusting the message-splitting ratios, offering a superior balance between interference management and decoding overhead.

Studies have corroborated its benefits, for example, in \cite{RS3}, an RSMA-based UAV communication system was proposed where joint optimization of scheduling, rate allocation, power control, and UAV trajectory was solved via a block coordinate descent approach, demonstrating superior energy efficiency and user fairness compared to OMA and NOMA schemes.
In \cite{RS4}, a low complexity user-pairing strategy was proposed for uplink RSMA to minimize maximum latency, achieving performance comparable to optimal RSMA while significantly reducing computational complexity compared to exhaustive search of the decoding order.
In \cite{RS5}, an online optimization framework using Lyapunov optimization and successive convex approximation was developed for RSMA-based UAV networks, effectively minimizing energy consumption while guaranteeing throughput requirements for mobile ground users through joint trajectory design and resource allocation.
As shown in  \cite{RS1}, RSMA-based PAS with multiple waveguides achieved higher sum rates compared to NOMA and conventional SDMA, thanks to its ability to reduce spatial correlation and enhance common stream rates. 
Furthermore, authors in \cite{RS2} derived closed-form outage probability expressions for uplink RSMA-PAS, validating its superiority over NOMA in terms of reliability and flexibility.

Despite these advancements, a significant limitation persists: existing multiple-access PAS frameworks are not content-aware. They cannot handle redundancy in user requests, leading to inefficient transmission of duplicate private streams and unnecessary interference. Consequently, this results in suboptimal latency performance, especially in multicast scenarios with popular content. Furthermore, the joint optimization of the pinching antenna's physical location and content-aware RSMA's resource allocation specifically for latency minimization remains an open challenge. 

\subsection{Our Contributions}
To bridge these gaps, this paper proposes a novel content-aware RSMA-enabled pinching-antenna  scheme that jointly optimizes antenna positioning and advanced content-aware RSMA's resource allocation to minimize average latency in multicast downlink systems. 
Unlike existing studies, our approach introduces a content-aware stream-sharing mechanism in RSMA where users requesting the same file share the same private stream, thereby reducing inter-user interference and power fragmentation. 
 The main contributions of this work are summarized as follows:
\begin{itemize}
    \item  We propose a novel content-aware RSMA-enabled downlink pinching-antenna framework, which integrates the spatial flexibility of the pinching antenna with a content-aware transmission strategy. 
    In the advanced content-aware RSMA strategy, users who request the same content share a unified private stream.
    This design dramatically reduces inter-user interference and improves power efficiency, achieving superior performance particularly when user requests are concentrated. 
    Furthermore, the private rate can be further improved by optimizing the position of the pinching antenna. For the common stream, all users need to decode it, so the common rate is limited by the user with the worst channel condition. By adjusting the position of the pinching antenna, the channel quality of the worst-case user can be improved, thereby enhancing the common rate.
    \item With the novel framework, we formulate the average latency minimization problem via a joint pinching antenna's location and advanced RSMA resource allocation perspective. The formulated problem is a nonconvex problem, and all variables are mutually coupled, which is difficult to solve directly.
    \item  To make it tractable, we develop a Content-Aware RSMA and Pinching-antenna Joint Optimization (CARP-JO) algorithm, which decomposes the original problem into two subproblems:  the advanced RSMA resource allocation subproblem and the antenna positioning subproblem.
    \item For the advanced RSMA resource allocation subproblem, we optimize power and rate by employing alternating optimization between private power control via bisection and common rate allocation via convex programming, coupled with a one-dimensional search over the common stream power. For the antenna positioning subproblem, we determine the optimal location through a golden-section search along the waveguide, solving a convex rate allocation problem at each candidate position to evaluate the system latency.
\end{itemize}
Extensive simulation results validate that our scheme significantly outperforms Traditional RSMA, NOMA, and Fixed antenna systems in various network scenarios, including changes in transmit power, user density, and coverage size.

The remainder of this article is structured as follows: Section II introduces the system model of the advanced RSMA-driven pinching antenna network under consideration, along with the principles of content transmission. Section III provides a detailed formulation of the problem, while Section IV presents the proposed efficient joint optimization framework. Extensive performance evaluations and comparisons are discussed in Section V. Finally, Section VI concludes the paper and outlines potential directions for future research. For brevity, the notations used throughout this paper are summarized in Table I. 
\begin{table}[h]
\centering
\caption{Notations and Descriptions}
\begin{tabular}{c|l}
\hline
 \multicolumn{1}{c|}{Notation} & \multicolumn{1}{c}{Description}\\
\hline
 $K$& The number of users\\
 $I$ & The number of contents \\
 $\bm{f}_k$ & The content request distribution for user\\
 $m_k$        &  The requested content state of user $k$\\
 $L_W$ & The length of dielectric waveguide\\
$D_x \times D_y$ & The side length of the user distribution area\\
$\Phi^\text{Pin} $ & The location of the pinching antenna \\
$\Phi_k$ & The location of user $k$ \\
$h_k$ & The channel between the antenna and user $k$ \\
$c$ &The speed of light\\
$f_c$  & The carrier frequency\\
$\cI(\bm{m})$ & The collection of distinct files \\
&under request state $\bm{m}$\\
$\cK_i$ & The user cohort requesting file $i$\\
$P_i$ & The  power for private part of content $i$\\
$P_0$ & The common stream power\\
$h_0$ & The worst-case channel gain\\
$R_0$ & The common stream rate\\
$r_i$ & The common rate allocation to content $i$\\
$R_{i,r}$ & The rate of the private stream $s_i$\\
$R_i$ & The total achievable rate for content $i$\\
$c_i$ &  The data size of content $i$\\
$\text{L}_\text{RSMA}$ &   Overall transmission latency for all users \\
$\bP$ &  The RSMA power control policy \\
$\br$ & The common rate allocation vector\\
\hline
\end{tabular}
\label{tab:simulation_parameters}
\end{table}

\section{System Model} \label{systemmodel}
We investigate a downlink PAS  where a base station (BS) serves a set of \(K\) users, as illustrated in Fig. 1. The BS employs a single dielectric waveguide of length \(L_W\) for transmission, with a pinching antenna activated at a specific point along it to establish LoS connections with the users, denoted by the set \(\mathcal{K} = \{1, \dots, K\}\). The system operates in a 3D Cartesian coordinate system: the waveguide is aligned along the \(x\)-axis at a fixed height \(d\), while the users are randomly distributed within a rectangular service area of dimensions \(D_x \times D_y\) in the \(x\)–\(y\) plane. The location of the pinching antenna is \(\Phi^{\text{Pin}} = (x^{\text{Pin}}, 0, d)\), and the location of user \(k\) is \(\Phi_k = (x_k, y_k, 0)\), where \(x_k \in [0, D_x]\) and \(y_k \in [-D_y/2, D_y/2]\) for all \(k \in \mathcal{K}\).

\begin{figure}
    \centering
    \includegraphics[width=1\linewidth]{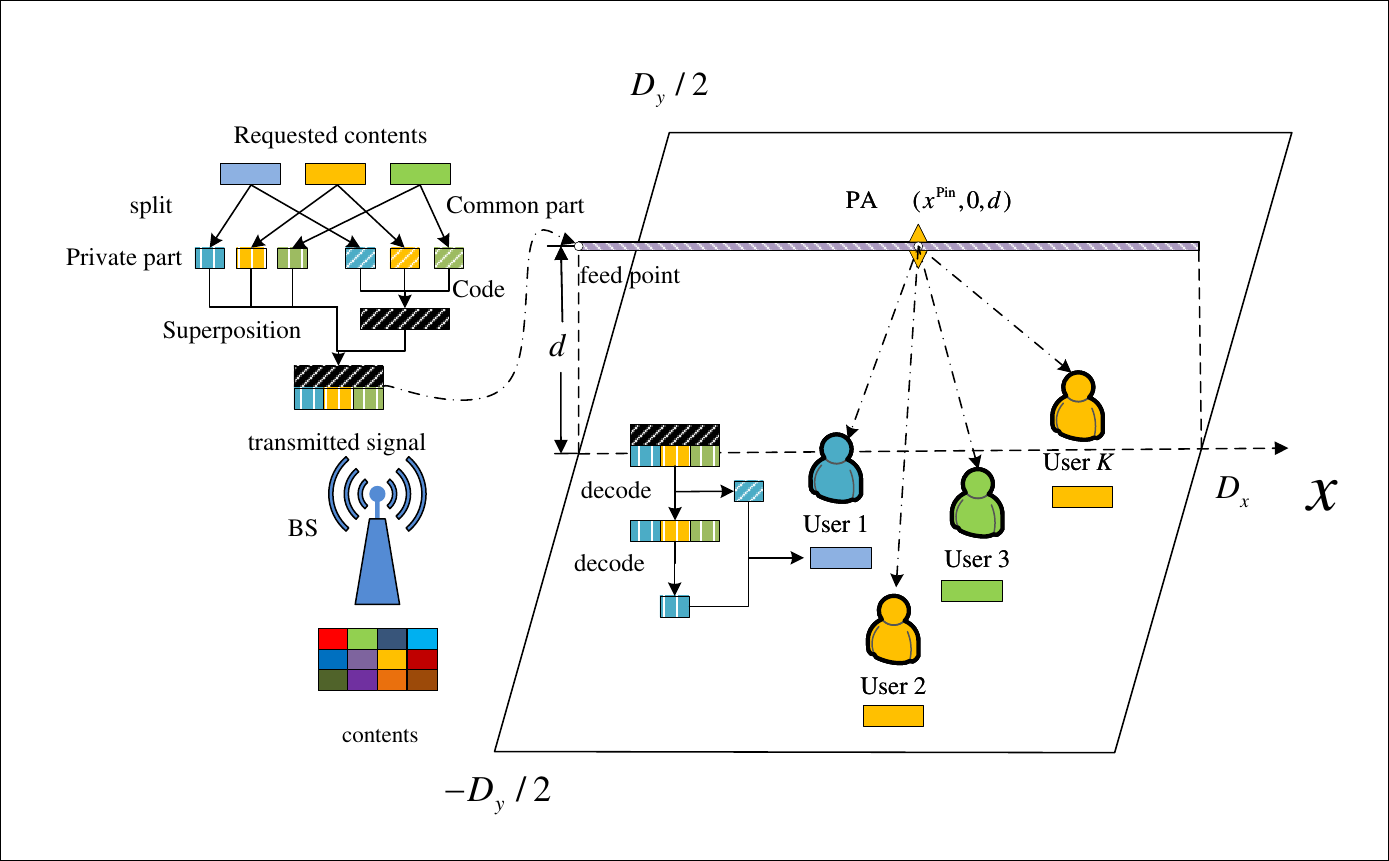}
    \caption{ System model of the considered content-aware RSMA-enabled PAS.
    }
    \label{sys}
\end{figure}

\subsection{Content Request Model}
The BS maintains a content library \(\mathcal{I} = \{1, 2, \dots, I\}\), consisting of \(I\) distinct files. Each content \(i \in \mathcal{I}\) is characterized by a data size \(c_i > 0\).  The content request distribution for user $k$ is represented by the vector $\bm{f}_k = \big( f_k(1), f_k(2), \ldots, f_k(I) \big)$, which follows a Zipf distribution \cite{zipf} and $\sum_{i=1}^If_k(i)=1$. 
Let \(m_k \in \mathcal{I}\) denote the content requested by user \(k\), with the request probability given by $f_k(m_k), ~m_k \in \mathcal{I}$.
User requests are assumed to be independent, thus the joint request state of the system is defined as \(\bm{m} = (m_k)_{k \in \mathcal{K}} \in \mathcal{I}^K\). 
Let $f(\bm{m})$ be the probability mass function of the joint request state $\bm{m}$, which is given by
\begin{equation} \label{psys}
f(\bm{m})  = \prod_{k \in \mathcal{K}} f_{k}(m_k).
\end{equation}

\subsection{Channel Model}
The channel between the BS and users in the PAS is characterized by a unique two-hop transmission process: the signal first propagates along the dielectric waveguide from BS to PA and is then radiated to the users via PA.
Let $\Phi^\text{BS}=(0,0,d)$ be the location of the feed point on BS. The signal fed into the waveguide at the feed point $\Phi^\text{BS}$ propagates along the waveguide to PA $\Phi^{\text{Pin}}$. This process incurs a phase shift, which is modeled by the complex gain between the feed point $\Phi^\text{BS}$ and PA $ \Phi^{\text{Pin}}$ \cite{PA14}:
\begin{equation}
h_{\uparrow} = e^{-j\frac{2\pi}{\lambda_g}\|\Phi^\text{BS} - \Phi^{\text{Pin}}\|},
\end{equation}
where \( \lambda_g = \lambda / n_{\text{eff}} \) is the guided wavelength, \( n_{\text{eff}} \) is the effective refractive index of the dielectric waveguide, and \( \|\Phi^\text{BS} - \Phi^{\text{Pin}}\| \) is the propagation distance within the waveguide.
The PA then radiates the signal into the free space, the channel between the PA at \( \Phi^{\text{Pin}} \) and user \( k \) at \( \Phi_k \) is accurately characterized by the free-space path loss model \cite{PA14}:
\begin{equation}
h_{\downarrow, k} = \frac{\sqrt{\eta} \, e^{-j\frac{2\pi}{\lambda}\|\Phi^{\text{Pin}} - \Phi_k\|}}{\|\Phi^{\text{Pin}} - \Phi_k\|},
\end{equation}
where \( \eta = \lambda^2/(16\pi^2) \) is the channel power gain at a reference distance of 1 m.

The overall channel gain for user \( k \) is the product of the waveguide propagation gain and the free-space radiation gain. Considering that the phase terms are compensated or averaged out in power calculations for system-level optimization \cite{PA14}, the effective channel gain of user $k$ used for resource allocation, referred to as $h_k$, is expressed as follows:
\begin{equation}
h_k = |h_{\uparrow} \cdot h_{\downarrow, k}|^2 = \frac{\eta}{\|\Phi^{\text{Pin}} - \Phi_k\|^2}.
\end{equation}
This simplified yet accurate model is justified by the PAS's operational principle, the dynamic positioning of the PA ensures a strong and stable LoS path, making the received power predominantly determined by the distance-dependent path loss. This model provides a tractable foundation for the subsequent joint optimization of antenna positioning and RSMA parameters.

\subsection{Advanced RSMA Transmission Model}
In traditional RSMA implementations, each user's message undergoes a segmentation process, splitting into common and private components. The common segments from all users undergo superposition coding to form a unified common stream $s_0$, decoded via a codebook shared across all receivers. Meanwhile, each private segment is independently encoded into a dedicated private stream per user.
Crucially, even when multiple users request identical content, each receiver still obtains a distinct private stream. This approach inherently generates substantial inter-user interference. Furthermore, given the base station's finite transmit power, replicating private streams for identical content requests fragments the available power resources. Consequently, the transmission power allocated to individual users becomes critically low.

Mitigating private stream interference is achieved by having users requesting identical content share a common private part. For a given request state $\bm{m}$, define $\mathcal{I}(\bm{m})$ as the collection of distinct files requested. Each content $i \in \mathcal{I}(\bm{m})$ has its private segment multicast via a dedicated private stream $s_i$.
Let $\mathcal{K}_i$ denote the user cohort that requests file $i$, implying that these users share the stream $s_i$. The transmit power allocated to content $i$ is denoted $P_i$, and $P_0$ is the transmit power allocated to common stream $s_0$. Crucially, the number of distinct requested files cannot exceed the user count: $|\cI(\bm{m})| \leq K$
where $|\mathcal{I}(\bm{m})|$ counts unique requested files. 
The transmitted signal is:
\begin{equation}
  s = \sqrt{P_0} s_0 + \sum_{i \in \mathcal{I}(\bm{m})} \sqrt{P_i} s_i.  
\end{equation}
This structure substantially reduces private stream transmissions in advanced RSMA. Under BS power constraints, concentrating power allocation yields higher per-stream transmission power, consequently enhancing achievable rates.

\subsubsection{Common Stream Rate}
Universal decoding of the common stream \(s_0\) mandates adherence to the limiting user channel. The worst-case channel gain \(h_0 = \min_{k\in \cK}h_k\) governs the common stream rate \(R_0\), which must satisfy:
\begin{align}
R_0=B\log_2(1+\frac{P_0 }{\sum_{i \in \cI(\bm{m})} P_i+ \frac{\sigma^2}{h_0}}),
\end{align}
where $B$ is the bandwidth, \(P_0\) designates the common stream power and \(P_i\) signifies the private stream power for content \(i\) (per prior definition). In addition,  $\sigma^2$ expresses the power of additive white Gaussian noise (AWGN). The common rate \(R_0\) is allocated to the different contents. Let \(r_i\) denote the common rate allocated to content \(i \in \mathcal{I}(\bm{m})\). These allocated rates \(r_i\) are non-negative and must satisfy the following sum constraint:
\begin{align}
\sum_{i \in \cI(\bm{m})} r_i\leq R_0,~r_i\geq 0.
\end{align}

\subsubsection{Private Stream Rate}
After successfully decoding the common stream $s_0$, each user proceeds to decode the private stream associated with its requested content. The rate of the private stream $s_i$ is denoted by $R_{i,r}$. To guarantee successful decoding of $s_i$ by all users in $\cK_i$, the rate $R_{i,r}$ must satisfy the following expression:
\begin{align}
R_{i,r}=B\log_2(1+\frac{P_i }{\sum_{j \in \cI(\bm{m}), j \neq i} P_j+ \frac{\sigma^2}{\min_{k\in \cK_i}h_k}}).
\end{align}
While serving users with poor channel conditions may appear to diminish achievable rates, the advanced RSMA scheme enhances private stream rates by reducing the number of distinct private streams, thereby concentrating available transmit power on each remaining stream.

\subsubsection{SIC Feasibility Condition}
To ensure successful successive interference cancellation (SIC) \cite{ref1}, the following condition must be satisfied:
\begin{equation}
P_0 \geq \frac{P_b}{2}+\frac{\sigma^2+\theta}{h_0},
\end{equation}
where $\theta$ denotes the minimum required power difference between the common stream power and the combined power of residual signals and noise. Additionally, $P_b$ corresponds to the total power budget available at the BS.

\subsection{Latency Model}
Based on the above analysis, the total achievable rate $R_i$ for content $i$ comprises both the common and private stream rates, and is obtained as:
\begin{equation}
R_i=r_i+R_{i,r}, ~ i \in \cI(\bm{m}).
\end{equation}
This work aims to minimize the average latency of the system. Denote by $\text{L}_\text{RSMA}$ the overall transmission latency for all users. The system latency under each request state is determined by the maximum latency among all requested contents, which is defined as:
\begin{equation}
\text{L}_\text{RSMA}=\max_{i \in \cI(\bm{m})} \frac{c_i}{R_i}. \label{tr}
\end{equation}

\section{Problem Formulation and Analysis} \label{problem}
This section details the mathematical formulation of the latency minimization problem and analyzes its inherent challenges. First, we formally present the joint optimization problem that integrates pinching-antenna positioning with content-aware RSMA resource allocation. Subsequently, we analyze the key difficulties in solving this problem, including its non-convexity and the strong coupling among variables, which motivates the design of our proposed alternating optimization framework.
\subsection{Problem Formulation}
This paper aims to minimize the average latency in content-aware RSMA-enabled pinching-antenna networks by jointly optimizing antenna positioning and advanced RSMA resource allocation strategies.
For notational simplicity, let $\bP = (P_i)_{i \in \cI(\bm{m}) \cup {0}}$ represent the RSMA power control policy, and $\br = (r_i)_{i \in \cI(\bm{m})}$ denote the common rate allocation vector.
Notably, the content-oriented RSMA resource allocation and the pinching-antenna location are adaptive to the request state $\bm{m}$. Therefore, we express these dependencies as $\bP(\bm{m})$, $\br(\bm{m})$, and $x^\text{Pin}(\bm{m})$.
The average system latency, denoted by ${\text{L}}(x^\text{Pin},\bP,\br)$, is consequently calculated as:
\begin{align} \label{objt}
{\text{L}}(x^\text{Pin},\bP,\br)=&\bbE[\text{L}_\text{RSMA}(x^\text{Pin}(\bm{m}),\bP(\bm{m}),\br(\bm{m}))], \notag\\
=&\sum_{\bm{m} \in \bm{M}}f(\bm{m}) \cdot \max_{i \in \cI(\bm{m})}  \frac{c_i}{R_i},
\end{align}
where $\bm{M}$ denotes the set of all possible cases of $\bm{m}$, $\bbE$ represents the expectation operator over all possible request states within $\bm{M} \in \cI^K$.

Based on the above definitions, the average latency minimization problem is formally established as follows\footnote{Note that all variables inherently depend on $\bm{m}$; the simplified notation is adopted for clarity.}:
\begin{subequations} \label{eq:main_optimization}
\begin{align}
\textbf{P0}:~&\mathop{\min}_{x^\text{Pin},\bP,\br} {\text{L}}(x^\text{Pin},\bP,\br)   \notag \\
\text{s.t.}~& \sum_{i \in \cI(\bm{m})}P_i +P_0 \leq P_b, ~ \bm{m} \in \bm{M}, \label{C1} \\
& \sum_{i \in \cI(\bm{m})} r_i\leq R_0, ~ \bm{m} \in \bm{M}, \label{C2} \\
& P_0 \geq \frac{P_b}{2}+\frac{\sigma^2+\theta}{h_0}, \label{C3} \\
& x^\text{Pin} \in [0,L], \label{C4} \\
& P_i \geq 0, \quad i \in \cI(\bm{m}) \cup \{0\}, ~\bm{m} \in \bm{M}, \label{C5} \\
& r_i\geq 0, \quad i \in \cI(\bm{m}), ~\bm{m} \in \bm{M}, \label{C6}
\end{align}
\end{subequations}
where constraint  \eqref{C1} imposes the total power budget limitation at the BS.  \eqref{C2} restricts the aggregate common rate allocation. \eqref{C3} ensures the feasibility of SIC. \eqref{C4} confines the antenna position to the length of the dielectric waveguide. Constraints \eqref{C5} and \eqref{C6} impose non-negativity requirements on the corresponding decision variables.

\subsection{Problem Analysis}
\subsubsection{Challenges} The formulated average latency minimization  \textbf{P0} is challenging to solve directly due to its inherent non-convexity and the strong coupling among the optimization variables. Specifically, the following challenges are identified: \textit{i) Non-convexity and NP-hardness.} The objective function involves a max-of-ratio structure over multiple users and content items, while the constraints include non-convex expressions in both the rate and power variables. This makes \textbf{P0} a non-convex optimization problem, which is generally NP-hard. \textit{ii) Strong coupling among variables.} The pinching antenna position 
\( x^{\text{Pin}} \) affects all user channel gains, thereby coupling the physical layer configuration with the higher-layer resource allocation. Meanwhile, the RSMA power and rate variables are mutually interdependent, making it difficult to optimize them separately. \textit{iii) Extensive and mixed decision space.} The problem involves both continuous variables (e.g., power, rate, antenna position) and discrete-like dependencies (e.g., content-aware stream grouping), leading to a complex mixed-integer non-linear programming (MINLP) structure.
\subsubsection{Motivation for the Proposed Approach} The aforementioned challenges stimulate the design of a low-complexity and high-efficiency algorithm. Our proposed solution, an Alternating Optimization (AO) framework, is guided by the following key motivations:
\begin{enumerate}[label=\roman*)]  
    \item \textit{Tackling NP-hardness and Non-convexity.} The NP-hard and non-convex nature of \textbf{P0} makes finding a global optimum computationally prohibitive. While machine learning algorithms like deep reinforcement learning (DRL) are alternatives, the mixed and coupled decision spaces in \textbf{P0} would lead to complex action spaces, requiring extensive environmental interactions and suffering from long convergence times. In contrast, our proposed AO framework decomposes the problem into tractable subproblems, enabling efficient and stable convergence to a high-quality suboptimal solution.

    \item \textit{Decoupling Interdependent Variables.} The tight coupling among optimization variables necessitates a co-design approach. However, solving them simultaneously is intractable. This motivates us to decouple \textbf{P0} into manageable subproblems: the RSMA resource allocation subproblem for a fixed antenna position and the antenna positioning subproblem under given RSMA parameters. This decomposition effectively breaks the coupling, allowing us to leverage powerful convex optimization and one-dimensional search techniques for each subproblem.
    \item \textit{Balancing Complexity and Performance.} When decomposing the problem, the extensive decision space poses a challenge in trading off complexity reduction and performance degradation. Dividing the problem into more subproblems lowers computational complexity but may narrow the solution space. Our decomposition strategy is driven by the need to reduce problem complexity while maintaining satisfactory performance. Specifically, we retain the joint optimization of power and rate within the RSMA subproblem due to their strong interaction, while separately optimizing the antenna position, which operates on a different spatial domain. This structure captures the most critical couplings while ensuring computational efficiency.
\end{enumerate}

\section{Algorithms Design} \label{algorithm}

This section presents the Content-Aware RSMA and Pinching-antenna Joint Optimization (CARP-JO) algorithm to solve the formulated \textbf{P0}. Owing to the non-convex nature of  $\text{L}_\text{RSMA}$ with respect to both and $x^\text{Pin}$,  \textbf{P0} is non-convex. As previously noted, all optimization variables are contingent upon the request state 
$\bm{m}$. Consequently, each distinct state necessitates independent variable optimization. To address this, we first consider a representative state 
$\bm{m} \in \bm{M}$, transform  \textbf{P0} into \textbf{P0'}, and solve it to obtain the minimal latency for that state. The overall average system latency is subsequently evaluated by aggregating results across all states according to equation \eqref{objt}. Therefore, in the following context, unless otherwise specified, we will focus on a specific state \(\bm{m}\). To make the subsequent optimization problems more concise, the constraint condition \(\bm{m} \in \bm{M}\) for \(\bm{m}\) will be omitted.
The  \textbf{P0'} is given as follows:
\begin{align*}
\textbf{P0'}: &\mathop{\min}_{x^\text{Pin},\bP,\br} \text{L}_\text{RSMA}    \\
\bm{\text{C1}}:~& \sum_{i \in \cI(\bm{m})}P_i +P_0 \leq P_b, \\
\bm{\text{C2}}:~& \sum_{i \in \cI(\bm{m})} r_i\leq R_0, \\
\bm{\text{C3}}:~& P_0 \geq \frac{P_b}{2}+\frac{\sigma^2+\theta}{h_0}, \\
\bm{\text{C4}}:~&         x^\text{Pin} \in [0,L], \\
\bm{\text{C5}}:~&  P_i \geq 0, i \in \cI(\bm{m}) \cup \{0\}, \\
\bm{\text{C6}}:~& r_i\geq 0, i \in \cI(\bm{m}).     
\end{align*}
It can be seen from constraint C2 that the three variables are mutually coupled, making it difficult to solve directly. Therefore, we split the  \textbf{P0'} into two subproblems: the resource allocation subproblem for RSMA and the pinching-antenna location optimization subproblem. 

\subsection{Resource Allocation of Advanced RSMA}
This subsection addresses the resource allocation subproblem for RSMA under a fixed antenna location \(x^\text{Pin}\), denoted as \textbf{P1}. The formulation is as follows:
\begin{align*}
\textbf{P1}:&\mathop{\min}_{\bP,\br} \max_{i \in \cI(\bm{m})}  \frac{c_i}{R_i}  \\
\text{s.t.}~ & \text{C1}, \text{C2}, \text{C3}, \text{C5}, \text{C6}. 
\end{align*}
To enhance tractability, an auxiliary variable \(\mu\) is introduced such that \(\mu \geq \frac{c_i}{R_i}\) for all \(i \in \cI(\bm{m})\). Then,  \textbf{P1} can be reformulated as:
\begin{align*}
\textbf{P1'}:&\mathop{\min}_{\bP,\br,\mu} \mu  \\
\text{s.t.}~ & \text{C1}, \text{C2}, \text{C3}, \text{C5}, \text{C6},\\
&\bm{\text{C7}}:~\frac{c_i}{R_i} \leq \mu,  i \in \cI(\bm{m}). 
\end{align*}
To efficiently solve \textbf{P1'}, we first initialize a feasible value for \(P_0\). With \(P_0\) fixed, we then perform alternating optimization between the common rate vector \(\br\) and the private power allocation vector \(\bP_r \triangleq (P_i)_{i \in \cI(\bm{m})}\). Finally, a one-dimensional search over \(P_0\) is conducted to determine its optimal value along with the corresponding optimal \(\br\) and \(\bP_r\).

\subsubsection{Private Power Control Algorithm}

Given a fixed common rate allocation vector \(\br\), the private power allocation subproblem is formulated as:
\begin{align*}
\textbf{P1.1}:&\mathop{\min}_{\bP_r,\mu} \mu  \\
\text{s.t.}~ & \text{C1}, \text{C5}, \text{C7}.
\end{align*}
Define \(N_i = \frac{\sigma^2}{\min_{k\in \cK_i}h_k}\) and \(P_\text{pri} = P_b - P_0\). For $i\in \cI(\bm{m})$,  constraint C7 can be rewritten as:
\begin{align}
&R_i \geq \frac{c_i}{\mu}, \\
&B\log_2\left(1+\frac{P_i}{\sum_{j \in \cI(\bm{m}), j \neq i} P_j+ N_i}\right) \geq \frac{c_i}{\mu} - r_i, \\
&B\log_2\left(\frac{P_\text{pri}+N_i}{P_\text{pri}+N_i-P_i}\right) \geq \frac{c_i}{\mu} - r_i.
\end{align}
Let \(\Psi_i = P_\text{pri} + N_i\), then constraint C7 becomes:
\begin{align}
\bm{\text{C7}'}:~B\log_2(\Psi_i - P_i) + \frac{c_i}{\mu} \leq B\log_2(\Psi_i) + r_i,~i\in \cI(\bm{m}).
\end{align}
Thus,  \textbf{P1.1} is reformulated as:
\begin{align*}
\textbf{P1.1'}:&\mathop{\min}_{\bP_r,\mu} \mu    \\
\text{s.t.}~ & \text{C1}, \text{C5}, \text{C7}'.
\end{align*}
\textbf{P1.1'} is non-convex due to the structure of constraint \(\text{C7}'\). Although the objective is linear in \(\mu\) and hence convex, the constraint involves the sum of a concave term \(B\log_2(\Psi_i - P_i)\) and a convex term \(c_i/\mu\), which together do not yield a convex feasible region.
Nevertheless, the problem can be efficiently solved via the bisection method by exploiting the monotonicity of \(\mu\). For a fixed \(\mu\), constraint \(\text{C7}'\) simplifies to:
\begin{equation}
\bm{\text{C7.1}}:~ P_i \geq \Psi_i \left(1 - 2^{\frac{r_i}{B} - \frac{c_i}{\mu B}}\right),~i\in \cI(\bm{m}).
\end{equation}
The resulting problem for fixed \(\mu\) becomes:
\begin{align*}
\textbf{P1.1}^\dagger:&\mathop{\min}_{\bP_r,\mu} \mu   \\
\text{s.t.}~ & \text{C1}, \text{C5}, \text{C7.1}.
\end{align*}

To apply bisection, the upper and lower bounds of \(\mu\) must be established. A natural lower bound is \(\mu_{\text{low}} = 0\). The upper bound \(\mu_{\text{high}}\) is derived from constraint C7 under the worst-case scenario where \(P_i = 0\), i.e., $\mu_{\text{high}} = \max_{i \in \cI(\bm{m})} \frac{c_i}{r_i}$.
The bisection procedure is then carried out as follows: at each iteration, compute the midpoint \(\mu_{\text{mid}} = (\mu_{\text{low}} + \mu_{\text{high}})/2\), and check the feasibility of  \textbf{P1.1}$^\dagger$. If feasible, set \(\mu_{\text{high}} = \mu_{\text{mid}}\); otherwise, set \(\mu_{\text{low}} = \mu_{\text{mid}}\). The process continues until convergence.  For clarity, the detailed procedure is summarized in Algorithm \ref{ppca}.

\begin{algorithm}[ht]
	\caption{Joint Optimization Algorithm for \textbf{P1.1}}
	\label{ppca}	
	\small
	\begin{algorithmic}[1]
		\REQUIRE Initial rate allocation $\br$ and the common stream power $P_0$
		\STATE Set \(\mu_{\text{low}} = 0\), $\mu_{\text{high}} = \max_{i \in \cI(\bm{m})} \frac{c_i}{r_i}$.
		\REPEAT
		\STATE Let \(\mu_{\text{mid}} = (\mu_{\text{low}} + \mu_{\text{high}})/2\),
        check the feasibility of  \textbf{P1.1}$^\dagger$ by CVX
		\STATE If feasible, set \(\mu_{\text{high}} = \mu_{\text{mid}}\), and record a feasible solution $\bP_r$
		\STATE otherwise, set \(\mu_{\text{low}} = \mu_{\text{mid}}\)
		\UNTIL{\(|\mu_{\text{low}} -\mu_{\text{high}}| \leq \epsilon_1\)}
		\RETURN Optimized variables $\bP_r$
	\end{algorithmic}
\end{algorithm}

\subsubsection{Rate Allocation Algorithm}

With the private power allocation vector \(\bP_r\) held fixed, the original constraint \(\text{C7}'\) can be rearranged into a more tractable form as follows:
\begin{equation}
    \bm{\text{C7.2}}:~ r_i - \frac{c_i}{\mu} \geq -R_{i,r}.
\end{equation}
This rearrangement leads to the formulation of the rate allocation subproblem, denoted as \(\mathbf{P1.2}\):
\begin{align*}
\textbf{P1.2}:&\mathop{\min}_{\br,\mu} \mu  \\
\text{s.t.}~ & \bm{\text{C2}}: \sum_{i \in \mathcal{I}(\bm{m})} r_i \leq R_0, \\
& \bm{\text{C6}}: r_i \geq 0, ~ i \in \mathcal{I}(\bm{m}), \\
& \bm{\text{C7.2}}: r_i - \frac{c_i}{\mu} \geq -R_{i,r}, ~ i \in \mathcal{I}(\bm{m}).
\end{align*}
A detailed convexity analysis of problem \(\mathbf{P1.2}\) reveals its well-structured nature. The objective function is linear in \(\mu\), and therefore convex. The constraint C2 is a linear inequality, while C6 consists of linear non-negativity constraints, both of which are convex. The key insight lies in analyzing the structure of constraint C7.2. The term \(\frac{c_i}{\mu}\) is convex in \(\mu\) for \(\mu > 0\) since \(c_i\) is a positive constant, making its second derivative with respect to \(\mu\) positive. Consequently, \(-\frac{c_i}{\mu}\) is concave. As \(r_i\) is linear, the entire left-hand side expression \(r_i - \frac{c_i}{\mu}\) is concave. A constraint where a concave function is greater than or equal to a constant (here \(-R_{i,r}\)) defines a convex feasible set.

Therefore, problem \(\mathbf{P1.2}\) minimizes a linear objective subject to convex constraints, confirming it is a convex optimization problem. This favorable structure guarantees that any locally optimal solution is also globally optimal and allows the problem to be efficiently and reliably solved to global optimality using standard convex optimization tools such as CVX.

\subsubsection{Joint Resource Allocation}

Under a fixed common stream power allocation \(P_0\), the joint optimization of the common rate vector \(\br\) and the private power allocation vector \(\bP_r\) is performed using an alternating optimization framework. This approach iteratively and alternately solves the two subproblems introduced previously, namely, the private power control  \textbf{P1.1} and the common rate allocation  \textbf{P1.2}. 
To promote convergence and enhance numerical stability, the solution obtained from one subproblem in each iteration is used as the initial value for the subsequent subproblem. This warm-start strategy effectively reduces the number of iterations required and helps avoid suboptimal local points.
After the alternating optimization process between \(\br\) and \(\bP_r\) converges for a given \(P_0\), a one-dimensional search is performed over \(P_0\) within its feasible domain to determine the value that minimizes the overall system latency. The corresponding optimal resource allocation, comprising both the power and rate vectors, is denoted as \((\bP^*, \br^*)\). This hierarchical optimization structure, inner alternating optimization nested within an outer one, dimensional search ensures efficient and effective resource coordination under the coupled nature of the system constraints.

\subsection{Pinching-Antenna Location Optimization}
Under a fixed power allocation policy, the optimization problem for determining the optimal pinching-antenna position and common rate allocation can be formulated as follows. It is important to note that the antenna position \(x^{\text{Pin}}\) directly influences the channel gain of the user with the worst channel condition, thereby affecting the achievable common rate \(R_0\). This dependency creates a coupling between the antenna position and the common rate allocation vector \(\br\), necessitating their joint optimization. The resulting optimization problem is formally stated as:
\begin{align*}
\textbf{P2}: &\mathop{\min}_{x^\text{Pin},\br} \max_{i \in \mathcal{I}(\bm{m})} \frac{c_i}{r_i + R_{i,r}(x^{\text{Pin}})}  \\
\bm{\text{C2}}:~& \sum_{i \in \mathcal{I}(\bm{m})} r_i \leq B\log_2\left(1 + \frac{P_0}{\sum_{i} P_i + \frac{\sigma^2}{h_0(x^{\text{Pin}})}}\right), \\
\bm{\text{C4}}:~& x^\text{Pin} \in [0,L],\\
\bm{\text{C6}}:~& r_i \geq 0, ~ i \in \mathcal{I}(\bm{m}).
\end{align*}
The direct solution of  \textbf{P2} presents significant challenges due to the strongly non-convex nature of the objective function and the intricate coupling between the optimization variables \(x^{\text{Pin}}\) and \(\br\). The objective function incorporates a max-min fractional structure, while the constraint \(\mathbf{C2}\) introduces a logarithmic relationship that depends nonlinearly on \(x^{\text{Pin}}\) through the worst-case channel gain \(h_0(x^{\text{Pin}})\).

However, a key observation enables an efficient solution approach: for any fixed antenna position \(x^{\text{Pin}}\), the problem reduces to a rate allocation subproblem that shares the same structural properties as the previously solved  \textbf{P1.2}. This suggests a decomposition strategy where we:
Leverage the one-dimensional nature of \(x^{\text{Pin}} \in [0,L]\) by employing the golden-section search method, which is particularly effective for optimizing unimodal functions in one-dimensional spaces, even when the objective function lacks closed-form differentiability.
For each candidate antenna position \(x^{\text{Pin}}\) evaluated during the golden-section search, solve the associated rate allocation subproblem:
\begin{align*}
\textbf{P2.1}:&\mathop{\min}_{\br} \max_{i \in \mathcal{I}(\bm{m})} \frac{c_i}{r_i + R_{i,r}(x^{\text{Pin}})}  \\
\bm{\text{C2}}:~& \sum_{i \in \mathcal{I}(\bm{m})} r_i \leq B\log_2\left(1 + \frac{P_0}{\sum_{i} P_i + \frac{\sigma^2}{h_0(x^{\text{Pin}})}}\right), \\
\bm{\text{C6}}:~& r_i \geq 0,~ i \in \mathcal{I}(\bm{m}).
\end{align*}
\textbf{P2.1} maintains a convex structure similar to  \textbf{P1.2}, as analyzed previously, and can therefore be efficiently solved to global optimality using established convex optimization techniques, such as those implemented in the CVX framework.

The complete algorithm proceeds as follows: the golden-section search iteratively narrows down the interval for \(x^{\text{Pin}}\) based on objective function evaluations. At each evaluation point, the corresponding convex rate allocation subproblem \textbf{P2.1} is solved to determine the achievable latency. This process continues until the search interval converges, ultimately identifying the optimal antenna position \(x^{\text{Pin}*}\) and the corresponding rate allocation vector \(\br^*\) that collectively minimize the system latency. This decomposition approach effectively handles the non-convexity of the original problem while ensuring computational efficiency.

\subsection{Joint Optimization Algorithm for  \textbf{P0}}

This subsection presents an alternating optimization algorithm designed to solve the original average latency minimization problem by jointly updating the variables \(x^\text{Pin}\), \(\br\), and \(\bP\). Let \(x^\text{Pin}(t)\), \(\br(t)\), and \(\bP(t)\) denote the antenna position, rate allocation, and power allocation in the \(t\)-th iteration, respectively. In addition, denote by $L(t)$ the average latency in the \(t\)-th iteration.
For the pecific request state $\bm{m}$,   the proposed algorithm proceeds as follows: given the antenna location from the previous iteration, \(x^\text{Pin}(t-1)\), we first compute \(\bP(t)\) and \(\br(t)\) by solving  \textbf{P1}, and obtain $L(t)$. Then, using the updated power allocation \(\bP(t)\), we jointly optimize the antenna position and rate allocation, yielding \(x^\text{Pin}(t)\) and \(\br(t)\) via  \textbf{P2}, and obtain $L'(t)$. This iterative process continues until the improvement in average latency falls below a predefined threshold \(\epsilon\). By applying the described alternating optimization procedure to each possible request state $\bm{m} \in \bm{M}$, we can obtain the corresponding optimal solution tuple $(x^\text{Pin}(\bm{m}), \bP(\bm{m}), \br(\bm{m}))$. Finally, the average system latency ${\text{L}}(x^\text{Pin},\bP,\br)$ is computed by substituting these state-dependent solutions into the expectation formula given in \eqref{objt}.
For clarity, the detailed procedure is summarized in Algorithm \ref{jotalg}.

\begin{algorithm}[ht]
	\caption{Joint Optimization Algorithm for \textbf{P0}}
	\label{jotalg}	
	\small
	\begin{algorithmic}[1]
		\REQUIRE Initial antenna location \(x^\text{Pin}(0)\) and optimal solution tuples for all states \(\bm{m} \in \bm{M}\): 
		
		\FOR{each request state \(\bm{m} \in \bm{M}\)}
		\STATE Set \(t = 1\)
		\STATE Set \(x^\text{Pin}(0) \gets x^\text{Pin}(0)\) for current state \(\bm{m}\)
		\REPEAT
		\STATE Update \(\bP(t)\) and \(\br(t)\) by solving  \textbf{P1} under fixed \(x^\text{Pin}(t-1)\)
		\STATE Compute average latency \(L(t)\) using \eqref{tr}
		\STATE Update \(x^\text{Pin}(t)\) and \(\br(t)\) by solving  \textbf{P2} under fixed \(\bP(t)\)
		\STATE Compute updated average latency \(L'(t)\) using \eqref{tr}
		\STATE \(t \leftarrow t + 1\)
		\UNTIL{\(|L(t) - L'(t)| \leq \epsilon\)}
		\STATE Store optimal solution for state \(\bm{m}\):
		\STATE \(x^\text{Pin}(\bm{m}) \gets x^\text{Pin}(t)\), \(\bP(\bm{m}) \gets \bP(t)\), \(\br(\bm{m}) \gets \br(t)\)
		\ENDFOR
		
		\STATE Compute the average system latency ${\text{L}}(x^\text{Pin},\bP,\br)$ by substituting these state-dependent solutions into the expectation formula given in \eqref{objt}
		\RETURN Optimal solution tuples \(\{(x^\text{Pin}(\bm{m}), \bP(\bm{m}), \br(\bm{m}))\}_{\bm{m} \in \bm{M}}\) and average latency ${\text{L}}(x^\text{Pin},\bP,\br)$
	\end{algorithmic}
\end{algorithm}

The proposed alternating optimization algorithm iteratively updates the antenna location, power allocation, and rate allocation variables, ensuring that the average system latency is monotonically non-increasing in each iteration. Since the objective function is bounded below and the constraint set is closed and convex, the algorithm is guaranteed to converge to a stationary point. 

\section{Simulation Results}
In this section, we conduct comprehensive numerical simulations to assess the performance of the proposed  CARP-JO algorithm against several baseline schemes. The simulation parameters are configured as follows \cite{PA5}: the carrier frequency is set to \(f_c = 28\) GHz, the antenna height is \(d = 3\) m, and the noise power is \(\sigma^2 = -90\) dBm. Four users are randomly distributed within a rectangular service area of dimensions \(D_x = 120\) m and \(D_y = 40\) m. These users request content from a catalog comprising 30 items, with content sizes uniformly distributed between 1 and 20 Mbits. The system bandwidth is assumed as 1 MHz. The popularity of the files follows a Zipf distribution with an exponent of 0.5. The BS transmit power budget is constrained to \(P_b = 25\) dBm, and the length of the dielectric waveguide is set to \(D_x\).
The simulation parameters are given in Table II.

\begin{table}[h]
\centering
\caption{Simulation Parameters}
\begin{tabular}{l|l}
\hline
 \multicolumn{1}{c|}{Parameters} & \multicolumn{1}{c}{Values}\\
\hline
The carrier frequency $f_c$& 28 GHz\\
The antenna height $d$  & 3 m\\
The noise power $\sigma^2$ & -90 dbm\\
The length $D_x$ and width $D_y$ of \\the distribution range of users & 120 m, 40 m\\
The number of contents & 30\\
The exponent of the Zipf distribution & 0.5 \\
The transmit power budget of BS $P_b$  & 25 dbm\\
\hline
\end{tabular}
\label{tab:simulation_parameters}
\end{table}

Performance comparisons are conducted against the following baseline methods:
\begin{itemize}
\item Traditional RSMA: This baseline employs the standard RSMA protocol \cite{ref1} for content delivery without the proposed enhancements.
\item NOMA: This approach substitutes RSMA with NOMA \cite{ref2}, while retaining the same caching strategy introduced in this work.
\item Fixed-antenna: In this scheme, the antenna is statically positioned at \((0, 0, d)\) meters\cite{PA5}, and the system employs the proposed advanced RSMA mechanism.
 \end{itemize}

\begin{figure}
    \centering
    \includegraphics[width=1\linewidth]{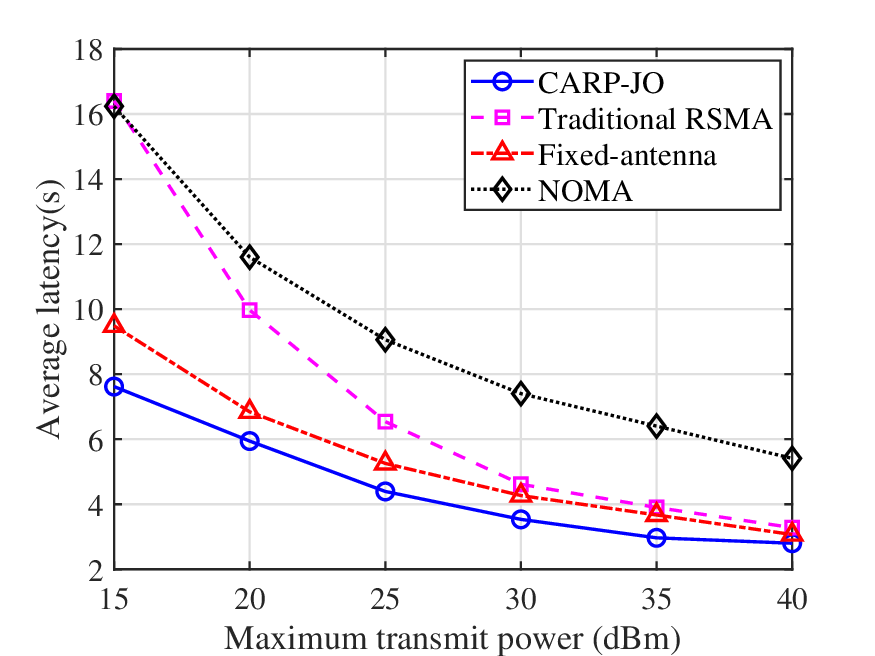}
    \caption{ System’s average latency vs. the maximum transmit
 power  of BS.}
    \label{fig1}
\end{figure}

Fig. \ref{fig1}  demonstrates the superior performance of the proposed CARP-JO scheme, as it consistently achieves the lowest average latency across the entire range of BS transmit power when compared to the three baseline methods. For instance, when the total power is 15 dbm, our CARP-JO scheme saves the latency by 20\%, 53\%, 54\% when compared to that of Fixed-Antenna, Traditional RSMA and NOMA, respectively.
This performance gain is attributed to the joint optimization of the antenna's physical location and the enhanced RSMA resource allocation, which dynamically minimizes path loss by establishing superior LoS links and efficiently manages interference and power through content-aware private stream sharing. 
All schemes benefit from increased transmit power, exhibiting lower latency due to improved SNR and higher data rates. However, the proposed CARP-JO  algorithm uniquely leverages the pinching antenna's flexibility to concentrate power and reduce spatial correlation. This enables it to significantly outperform conventional setups, particularly at low-to-moderate power levels.

\begin{figure}
    \centering
    \includegraphics[width=1\linewidth]{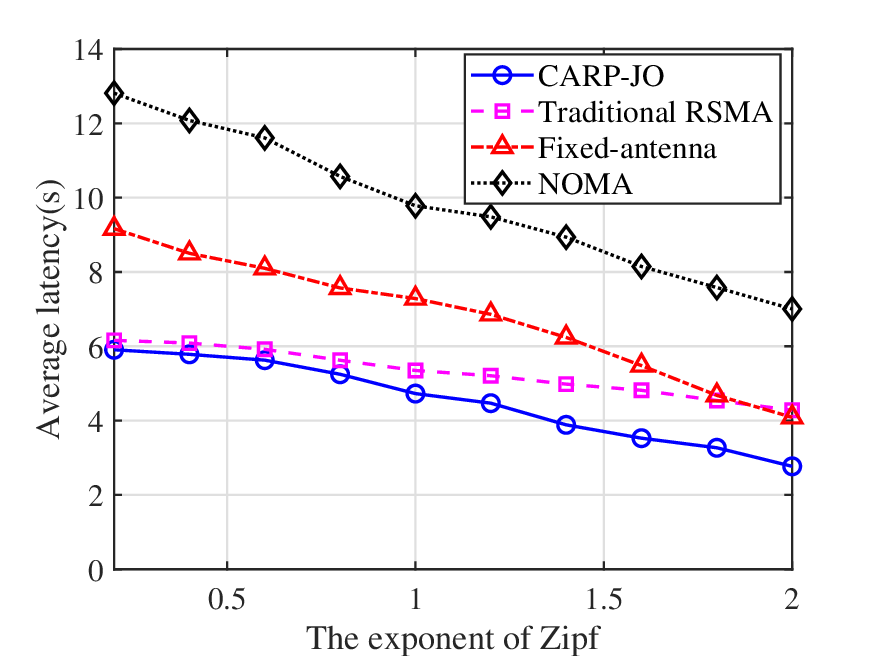}
    \caption{System’s average latency vs. the exponent of Zipf.}
    \label{fig2}
\end{figure}

Fig. \ref{fig2} illustrates the average latency performance with different exponents of Zipf. The proposed CARP-JO  scheme consistently achieves the lowest latency across the evaluated parameter range. Its advantages stem from the joint optimization of antenna positioning and its content-aware design, which is particularly effective when user requests are concentrated. A higher Zipf exponent leads to a more concentrated request pattern, increasing content redundancy; this scheme leverages that by having users requesting the same file share a private stream, drastically reducing interference and improving power efficiency. 
In contrast, the traditional RSMA scheme performs somewhat worse because it fails to exploit this content redundancy, inefficiently dedicating separate private streams to all users regardless of their requests, an approach whose inefficiency is exacerbated by a concentrated request pattern. 
The fixed-antenna scheme and the NOMA scheme perform even worse,
which highlights the significant individual contributions of both the dynamic antenna positioning and the enhanced RSMA structure to the overall latency reduction.

\begin{figure}
    \centering
    \includegraphics[width=1\linewidth]{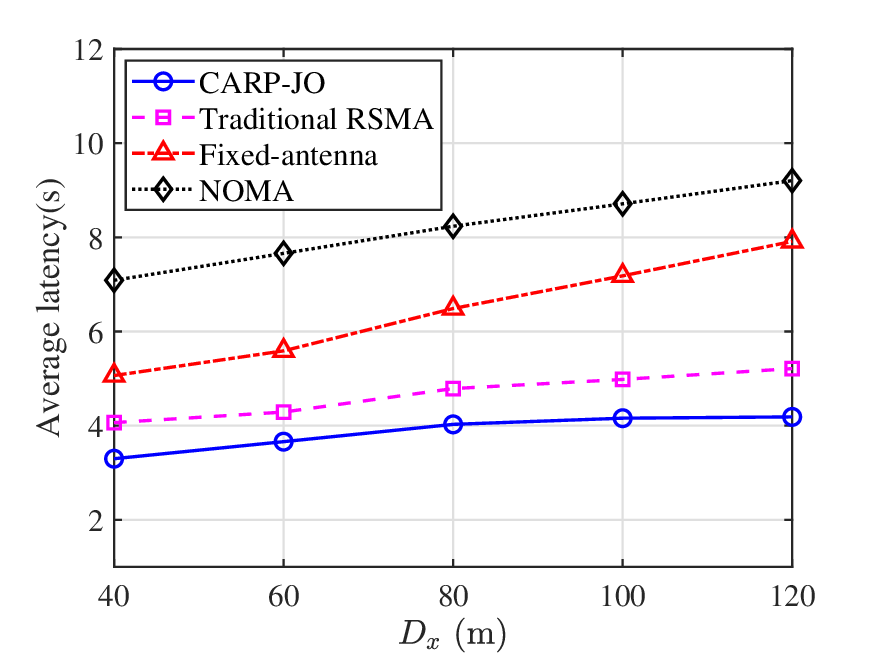}
    \caption{System’s average latency vs. $D_x$.}
    \label{fig3}
\end{figure}

Fig. \ref{fig3} illustrates the system's average latency as a function of the service area length, \(D_x\). The proposed CARP-JO scheme consistently achieves the lowest latency across the entire range of \(D_x\) values, demonstrating its robustness to changes in network coverage size. As \(D_x\) increases, the user distribution becomes more dispersed, leading to greater variation in path loss and a higher likelihood of poor channel conditions for edge users. The proposed CARP-JO scheme effectively mitigates this degradation through its dynamic antenna positioning, which optimizes the LoS link to the worst-case user, thereby improving the common stream rate \(R_0\). Concurrently, its content-aware design, which groups users requesting the same file onto a shared private stream, reduces inter-user interference and concentrates power, enhancing private rates \(R_{i,r}\). In contrast, the performance of the fixed-antenna scheme degrades more noticeably as \(D_x\) grows, highlighting the critical role of antenna mobility in large cells. Although Traditional RSMA and NOMA also use a movable antenna, their performance is still hampered by their core protocols. Traditional RSMA inefficiently allocates separate streams without exploiting content redundancy, while NOMA's rigid SIC structure struggles with the significant channel disparity in larger networks. Consequently, even with antenna mobility, they cannot match the holistic interference management and efficiency of the proposed co-designed scheme.

\begin{figure}
    \centering
    \includegraphics[width=1\linewidth]{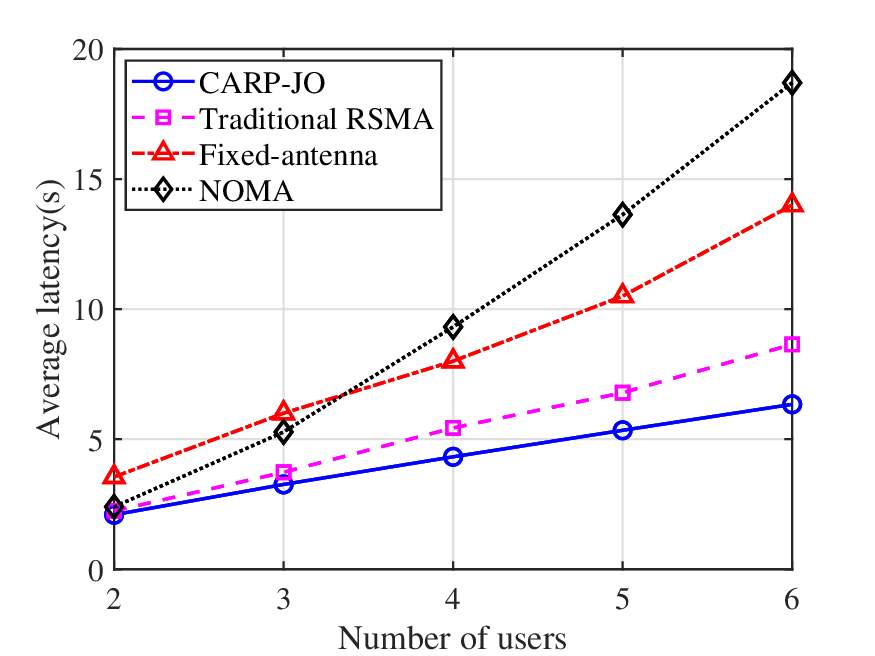}
    \caption{System’s average latency vs. the number of users}
    \label{fig4}
\end{figure}

Fig. \ref{fig4}  illustrates the system's average latency with different numbers of users.
It demonstrates that the proposed CARP-JO scheme maintains the lowest average latency as the number of users increases. 
When there are only two users, the performance of several algorithms is comparable, as the probability of users requesting the same file is lower when the number of users is small. Moreover, the fixed-antenna scheme exhibits higher latency compared to the pinching-antenna approach. 
As the number of users increases, the advantage of our proposed CARP-JO algorithm becomes evident. This is because, as the number of users increases, the likelihood of users requesting the same files rises, resulting in greater redundant interference among them, and our proposed CARP-JO  algorithm can effectively address this issue.
In contrast, the performance of Traditional RSMA and NOMA degrades more noticeably due to their inefficient handling of interference and lack of holistic optimization, while the fixed-antenna scheme is fundamentally limited by its inability to adapt to the changing user topology.

\begin{figure}
    \centering
    \includegraphics[width=1\linewidth]{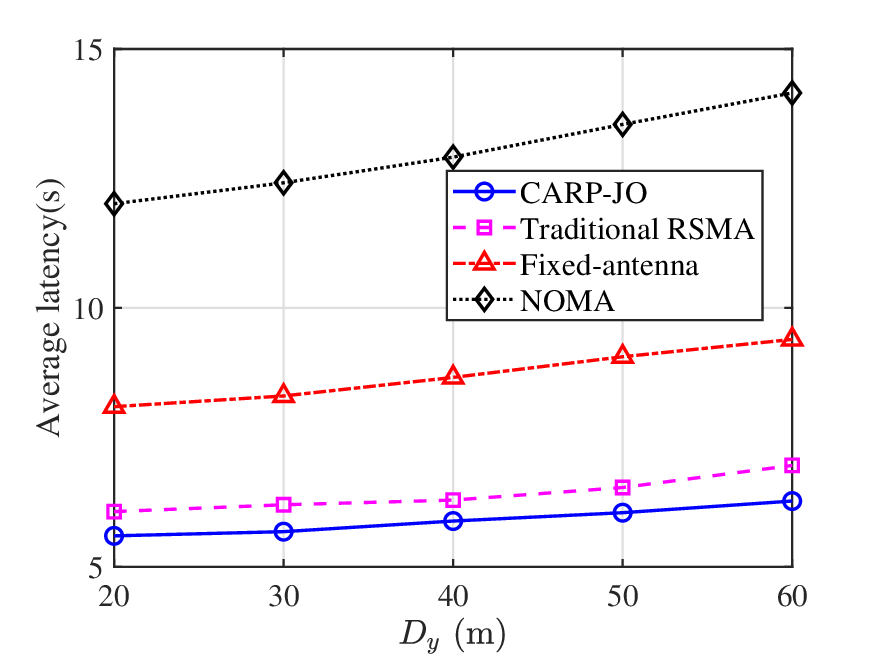}
    \caption{System’s average latency vs. $D_y$.}
    \label{fig5}
\end{figure}
Fig. \ref{fig5} illustrates the average latency performance with different width of user distribution area ($D_y$).
It shows that as $D_y$ increases from 40m to 120m, the average latency rises for all four schemes, with the proposed CARP-JO scheme consistently maintaining the lowest latency. While all methods experience performance degradation with greater distance, the proposed CARP-JO scheme demonstrates superior robustness, as its latency increases at a slower rate compared to Traditional RSMA, fixed-antenna configuration, and the significantly higher-latency NOMA baseline. When $D_y=60$m, the CARP-JO scheme saves the latency by 10\%, 33\%, 56\% when compared to that of Traditional RSMA, Fixed-Antenna and NOMA, respectively.
This indicates that the proposed CARP-JO  scheme is more effective in mitigating the adverse effects of increasing user distance, making it the most suitable for wide-area deployments.

\begin{figure}
    \centering
    \includegraphics[width=1\linewidth]{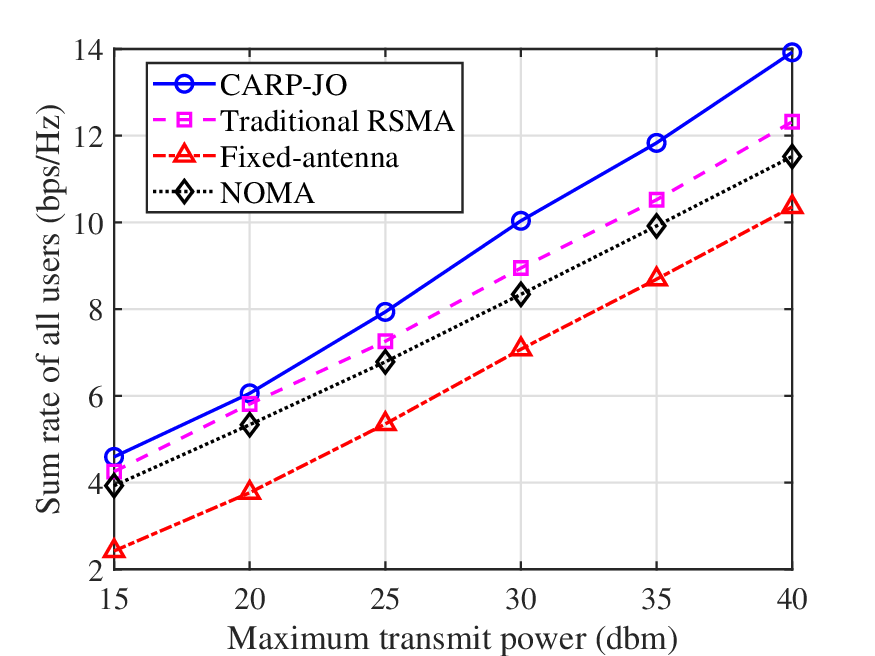}
    \caption{Sum rate of all users vs. the maximum transmit
 power  of BS.}
    \label{fig6}
\end{figure}
Fig. \ref{fig6} compares the sum rate of all users against the maximum transmit power for the proposed CARP-JO scheme and baseline methods. Notably, while NOMA exhibited the worst latency performance in our primary optimization, it surpasses the fixed-antenna baseline in terms of sum rate across most power levels. This occurs because the fixed-antenna system employs the proposed CARP-JO content-aware RSMA strategy, which is explicitly optimized for minimizing the maximum latency among all users. To achieve this, it prioritizes the user with the worst channel condition by allocating more resources, consequently limiting the rates achievable by users with better channels and thus reducing the overall sum rate. In contrast, the NOMA scheme, while not content-aware, benefits from the dynamic positioning of the pinching antenna, which improves channel conditions for all users. This allows NOMA to more effectively leverage its power-domain multiplexing to boost aggregate throughput, albeit at the cost of high latency and unfairness, as shown in our primary results.
Strikingly, the Proposed CARP-JO scheme transcends this trade-off. By synergistically co-designing the dynamic pinching antenna with the content-aware RSMA, it simultaneously achieves the highest sum rate and the lowest latency, demonstrating that intelligent physical layer and link layer integration can optimize both metrics concurrently.

\section{Conclusion}
This paper presented a novel content-aware RSMA-enabled pinching-antenna framework for downlink multicast systems, which integrates dynamic antenna positioning with content-sensitive transmission to minimize average latency. Within this architecture, we formulated a joint optimization problem that co-designs the physical placement of the pinching antenna and the RSMA resource allocation parameters to minimize system latency. Due to the non-convex and coupled structure of the problem, CARP-JO algorithm was developed, decomposing the original problem into tractable subproblems. RSMA resource allocation is addressed via a combination of bisection search and convex optimization, while antenna positioning is efficiently optimized using the golden-section search method. Extensive simulations validate that the proposed CARP-JO scheme consistently outperforms conventional RSMA, NOMA, and fixed-antenna systems under diverse scenarios, including variations in transmit power, user density, coverage size, and content popularity distribution. Future work will focus on extending multi-waveguide configurations, developing low-complexity algorithms for real-time applications, and ensuring robust operation under imperfect channel state information.

\bibliographystyle{IEEEtran}
\bibliography{ref}

\end{document}